\documentclass[a4paper,11pt]{article}
\pdfoutput=1 

\usepackage{jcappub} 
\usepackage[uncertainty-mode=separate,list-units=single]{siunitx}
\usepackage{soul,xcolor}
\usepackage[T1]{fontenc} 
\usepackage[shortcuts]{extdash}
\usepackage[normalem]{ulem}
\usepackage[nameinlink,noabbrev]{cleveref}
\usepackage{calc}
\usepackage{amsmath}
\usepackage{bm}
\usepackage{amssymb}
\usepackage{siunitx}
\usepackage{enumitem}
\usepackage{subcaption}
\usepackage{changes}
\usepackage{appendix}
\usepackage[numbers]{natbib}
\usepackage{cancel}
\crefname{equation}{eq.}{eqs.}
\Crefname{Equation}{Eq.}{Eqs.}

\DeclareSIUnit\steradian{steradian}

\providecommand{\sorthelp}[1]{}

\begin{document}

\title{HierArchical Wavelet Coefficients on the Sphere (HAWCS): the Scattering Transform on Spherical Maps}

\author[1,2]{Arefe Abghari,\note{Corresponding author.}}
\author[4]{Lukas T. Hergt,}
\author[2]{Douglas Scott,}
\author[2,3]{Raelyn M. Sullivan,}
\affiliation[2]{University of British Columbia, Vancouver, BC V6T1Z1, Canada}
\affiliation[3]{Institute of Theoretical Astrophysics, University of Oslo, Blindern, Oslo, Norway}
\affiliation[4]{Universit\'{e} Paris-Saclay, CNRS/IN2P3, IJCLab, 91405 Orsay, France}

\emailAdd{arefeabghari@phas.ubc.ca}
\emailAdd{lukas.hergt@ijclab.in2p3.fr}
\emailAdd{dscott@phas.ubc.ca}
\emailAdd{raelyn.sullivan@astro.uio.no}

\abstract{
Extracting Gaussian information from data is well understood, but characterizing non-Gaussianity is challenging.
We introduce \texttt{HAWCS}, a \texttt{healpy}-based Python package for efficiently computing wavelet scattering transform coefficients from full-sky maps. These coefficients provide compact summary statistics that are sensitive to higher-order structure and interactions across angular scales. We test the implementation on controlled Gaussian and non-Gaussian fields and demonstrate its computational efficiency.
We then apply \texttt{HAWCS} to thermal Sunyaev--Zeldovich maps derived from \textit{Planck} observations and four simulations, quantifying and comparing their higher-order statistical properties. 
We also use the method to expose the signatures of gravitational lensing in cosmic microwave background~(CMB) temperature maps. 
These results demonstrate that \texttt{HAWCS} is a practical tool for validating component-separation pipelines and cosmological simulations, and a promising summary statistic for characterizing non-Gaussian structures beyond the power spectrum.  These coefficients could also be used to constrain or generate simulated maps that reproduce the same statistical features of real data.
}

\maketitle

\section{Introduction}
\label{sec:intro}
Current and upcoming cosmological surveys, including LiteBIRD~\cite{LiteBIRD}, the Simons Observatory~\cite{SimonsObservatory2025}, the Square Kilometer Array (SKA)~\cite{Braun2015}, and \textit{Euclid}~\cite{Laureijs2011EuclidReport}, among others, will provide cosmological measurements over wide areas with unprecedented statistical precision. 
Extracting the full potential of these observations requires novel and advanced methods capable of handling large volume data sets, complex data products, and weak statistical signals. 
Additionally, the use of powerful simulations has become an essential tool for understanding and interpreting the data from cosmological surveys. These simulations are used to create mock data sets that mimic the statistical properties of the observed Universe, and can be used to test the performance of data analysis pipelines and to estimate the systematic errors and biases that may be present in the data. However, it is important to ensure that the simulations accurately reproduce the statistical properties of the underlying cosmological fields.  

In the case of Gaussian fields, the 2-point correlation function, or equivalently, the power spectrum in Fourier space, provides the full statistical information contained in the field. For non-Gaussian fields we would in principle need 3-point, 4-point and all higher level correlation functions for a complete statistical description. However, these functions become rapidly difficult to compute and analyse. Additionally, higher-order correlation functions are less intuitive and more complex to interpret than the 2-point correlation function, making it challenging to extract information from them. This has motivated the development of alternative summary statistics that capture particularly relevant non-Gaussian information more efficiently. Among these, the wavelet scattering transform (WST) has emerged as a promising multiscale representation that retains sensitivity to higher-order structure, while remaining computationally efficient and stable.

The scattering transform,\footnote{From a traditional physics and mathematics standpoint, the name ``scattering transform'' is hardly very descriptive! This method is neither a physical scattering process nor is it fully invertible (to be a transform), although there are similarities with both concepts. Information propagates through a branching cascade of wavelet coefficients, while losing its phase, a bit like a wave hitting a physical obstacle.} first introduced by Mallat in 2012~\cite{Mallat2012GroupScattering}, is based on a wavelet decomposition of the signal, which allows for the separation of different scales and structures in the data, followed by non-linear modulus operations. The resulting ``WST coefficients'' are locally translation invariant and stable to small deformations, while retaining information about higher-order interactions. The scattering transform is a convolutional network with fixed wavelet filters that shares architectural features with convolutional neural networks (CNNs). However, unlike CNNs, the scattering transform employs predefined wavelet filters rather than learned kernels, providing a mathematically well-founded representation with provable stability properties. These characteristics make it a promising statistical descriptor for non-Gaussian cosmological fields.  Whether it is useful in practice depends on whether it effectively captures the sort of non-Gaussian information often found in survey data -- we shall show that this is indeed the case.

WST's most useful applications are found in data-sparse, high-noise, scientific fields where interpretability is important. 
Since deep-learning models require massive data sets to learn pattern filters, fields like astrophysics or medical imaging struggle with them. WST bypasses this by using mathematically fixed wavelet filters to extract rich, non-Gaussian structural features instantly. 
In astrophysics, this method has been successfully used for comparing different interstellar medium simulations~\cite{Allys2019TheISM}. In cosmology, the WST has shown great promise in parameter estimation for surveys of weak lensing~\cite{Cheng2021WeakSensitivity,Cheng2020ATransform,marinichenko2025flamingobaryoniceffectsweak,Boone_2026,sui2025evaluatesufficiencycomplementaritysummary,zhou2025proposalconstructdarkmatteronlycounterpart}, 21-cm signals~\cite{Greig2022DetectingTransform,Sinha_2026}, dust map analysis~\cite{https://doi.org/10.48550/arxiv.2207.12527,shimabukuro2026waveletscatteringsignaturesfuzzydark}, and 3-D studies of large-scale structure fields~\cite{Valogiannis2021TowardsTransform,Valogiannis2022GoingTransforms}. In all of these studies the WST method was applied on Euclidean (flat) maps. 
However, many cosmological observables, including the cosmic microwave background, weak-lensing shear, and galaxy density fields, are naturally defined on the celestial sphere. Extending the scattering transform to spherical geometry is therefore essential for the analysis of current and upcoming cosmological surveys. Only a few investigations of the spherical formulation of WST have previously been applied in astrophysics and cosmology~\cite{McEwen_2007,McEwen_2007_1,Leistedt2013,mcewen2006directionalcontinuouswavelettransform,AbghariMSc,Mousset_2024}. 

In this work, we develop and validate a WST implementation for fields defined on the 2-sphere. We present a detailed description of the spherical wavelet and verify that our implementation preserves the key mathematical properties of the Euclidean scattering transform. We provide an open-source implementation, the HierArchical Wavelet Coefficients on the Sphere (\texttt{HAWCS}) package\footnote{\url{https://github.com/ArefeAbghari/scattering-transform-spherical/tree/master}}, based on \texttt{healpy}.
We describe our choice of spherical wavelets and the mathematical formalism of spherical WST in \cref{sec:morlet,section:formalism}.
We provide a number of scientific applications in \cref{section:results}. First, we compare WST coefficients computed from \textit{Planck} Sunyaev--Zeldovich (SZ) maps with those obtained from cosmological simulations. We show that, even when simulations reproduce broadly similar tSZ power spectra, they can exhibit substantially different higher-order, cross-scale structure. We also find appreciable differences between the MILCA and NILC \textit{Planck} reconstructions, demonstrating that the inferred non-Gaussianity is sensitive to component-separation choices and residual foreground contamination.
We also use this method for lensed CMB maps and show that the resulting scattering coefficients are consistent with the standard lensing convergence power spectrum $C_L ^{\kappa \kappa}$, providing a fast way of computing whethere there are lensing signatures in a CMB map.


\section{The Morlet Wavelet on the Sphere}
\label{sec:morlet}
Wavelets are chosen to be a family of localized band-pass filters. As we discuss later, this localization property is essential for capturing non-Gaussianity. In WST, the wavelets are generated by performing spatial and angular dilation on a mother wavelet: 
\begin{equation}
 \psi^{j, l}(\bm{n})=2^{-j} \psi\left(2^{-j} r_{l}^{-1} \hat{\bm{n}}\right),
\end{equation}
where $j$ characterizes the spatial scale, $l$ is the orientation of the wavelet, and $r_l$ is the rotation operator. Since the convolution of a spherical map with a directional (non-axisymmetric) wavelet produces a function that lives on the 3-dimensional rotation group (SO(3)), rather than on the sphere itself, we only use symmetric wavelets here, and we drop the orientation indices hereafter. 

The dyadic scaling factor $(2^{-j})$ produces a logarithmically spaced sequence of wavelets, with each successive scale approximately twice the spatial extent of the previous one. Although other scale progressions are possible, dyadic scaling is the standard choice in scattering transforms and multiresolution wavelet analysis. 
The normalization adopted here corresponds to an $L^1$
normalization of the wavelet family, which preserves the $L^1$
norm across scales. Other normalization conventions, including $L^2$ normalization, are also implemented in our software package.

For a 2-dimensional field, the Morlet wavelet is a suitable choice~\citep{Goupillaud1984,Mallat2012,Bruna2013}. It consists of an oscillatory wave localized by a Gaussian envelope and is therefore simultaneously localized in real and Fourier space. Its complex-valued structure allows the modulus operator to smoothly extract stable, non-oscillatory structural envelopes. Moreover, its clean Gaussian profile in the frequency domain prevents information leakage between transform layers while creating a tight frame that preserves total signal energy (the Littlewood-Paley condition) across different scales and rotations~\citep{Mallat2012,Bruna2013}.

Extending this construction to fields defined on the sphere requires accounting for the geometry of $\mathbb{S}^2$. Several approaches to spherical wavelets have been proposed, including constructions based on stereographic projection~\citep{Antoine1997WaveletsApproach,Antoine2002,Wiaux2005} and wavelets defined directly through a tiling of spherical harmonic space~\citep{Leistedt2013,McEwen2006}. We instead introduce an axisymmetric, real-space Morlet-like wavelet by replacing the Euclidean radial coordinate with the geodesic angular distance $\theta$ on the sphere. Explicitly we have
\begin{equation}
  \psi^j(\theta) = A e^{-(\theta^2/2\sigma^{j^2})}(e^{i\ell^j_0\theta}-B) ,
  \label{eq:morlet_sph}
\end{equation}
where $A$ and $B$ should be determined such that the wavelet is normalized and admissible (i.e., the zeroth mode vanishes.). 
We use the \texttt{kymatio}~\cite{andreux2022kymatioscatteringtransformspython} convention for $\sigma^j$ and $\ell^j_0$ by multiplying the angular size of each pixel, $a_0$ by $2^j$:

\begin{equation}
\begin{aligned}
\sigma^j &=0.8 \times 2^{j} a_0 ; \\
\ell^j_{0} &=\frac{3 \pi}{4 \times 2^{j} a_0}.
\label{eq:sigma_flat}
\end{aligned}
\end{equation}

\begin{figure}[htbp!]
\centering
  \begin{subfigure}[b]{0.38\textwidth}
    \centering
    \includegraphics[width=\textwidth]{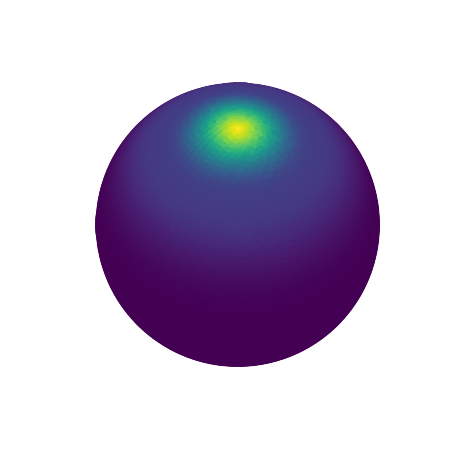}
  \end{subfigure}
\hfill
  \begin{subfigure}[b]{0.60\textwidth}
    \centering
    \includegraphics[width=\textwidth]{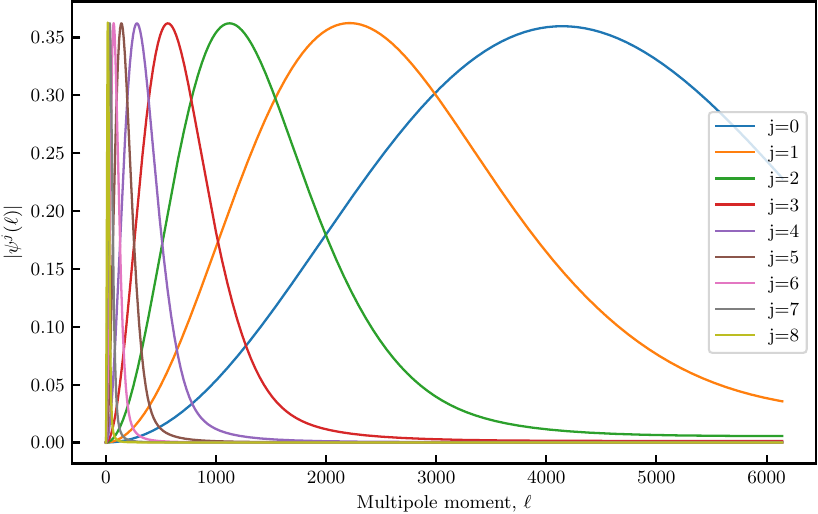}    
  \end{subfigure}
\caption{Absolute value of the axisymmetric spherical Morlet wavelet described by \cref{eq:morlet_sph}. {\it Left:} one wavelet profile in pixel space. {\it Right:} a set of wavelets with different scales in harmonic space. The $\ell = 0 $ mode is adjusted to zero to satisfy the admissibility condition. These wavelets are $L^1$ normalized.}
\label{fig:morlet_profile}
\end{figure}

This real-space construction provides an approximate spherical analogue of the Morlet wavelet. The quantities $\sigma^j$ and $\ell^j_0$ control the approximate central multipole and harmonic bandwidth. The construction does not guarantee an exact harmonic-space tiling or tight-frame condition, and curvature and pole-regularity effects may become significant for bigger wavelets. Nevertheless, the resulting spherical harmonic profiles are well localized and approximately Gaussian. As shown in \cref{fig:morlet_profile}, the filters provide distinct but overlapping coverage of the multipole range, while their vanishing monopoles ensure admissibility. The proposed wavelets therefore retain the band-pass localization and multiscale sensitivity required for the scattering transform, and, as we will show in the next section, they satisfy all useful properties of the scattering transform in Euclidean space.

\section{Mathematical Formalism of WST}
\label{section:formalism}
\begin{figure}[htbp!]
    \centering
    \includegraphics[width =1 \linewidth]{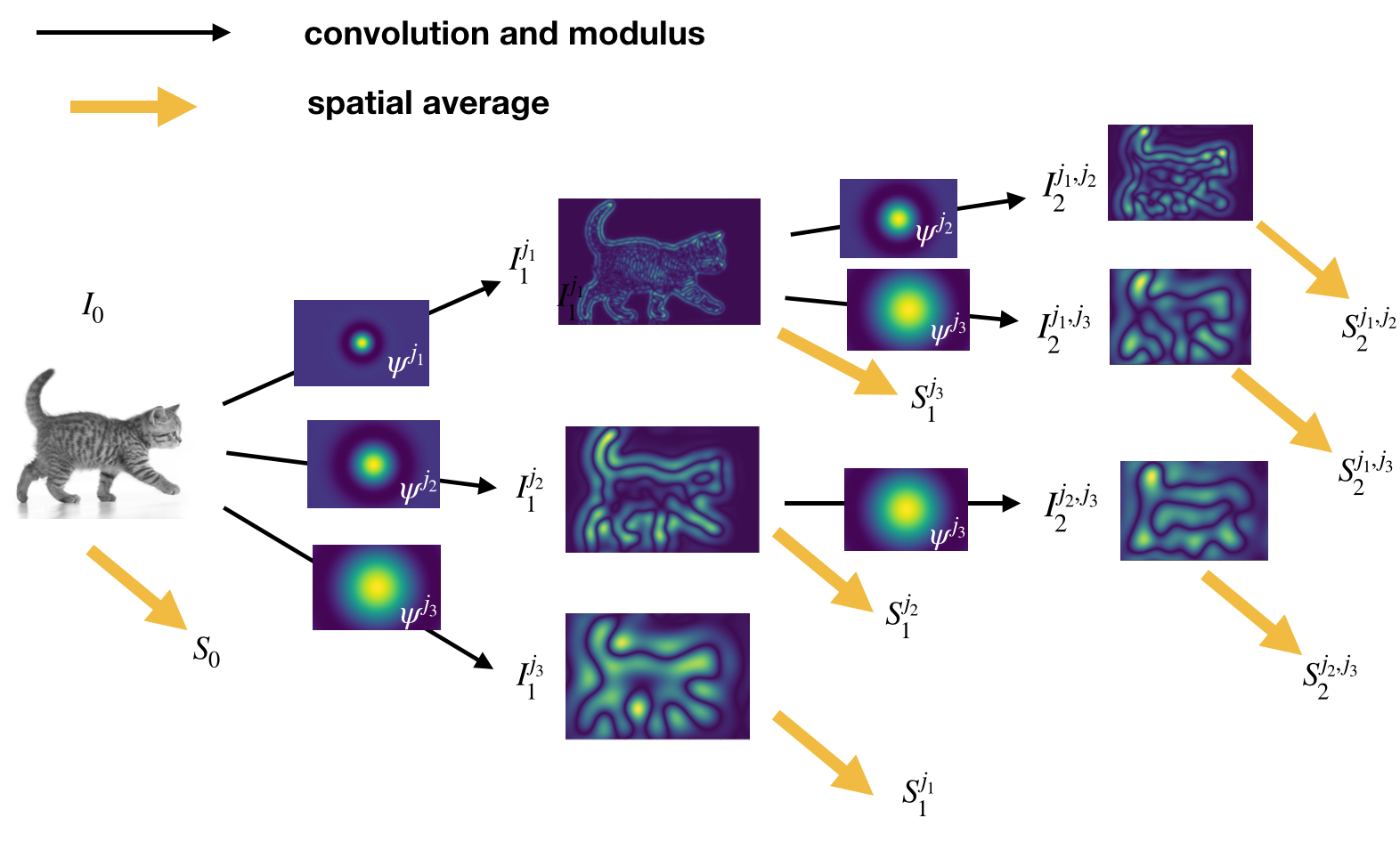}
    \caption{Schematic diagram showing the architecture of the WST algorithm up to second order. Each black arrow represents taking the convolution of the input field with the Morlet wavelet shown on the arrow, while the yellow arrow represents taking the spatial average. Here, we have used three sizes of wavelet. First-order convolutions pick up structures with the corresponding size from the input field. The $I_2$ fields pick up clusters of structures from the $I_1$ fields because each convolution smooths the field over the corresponding scale, our second convolution cannot have a smaller scale than the first one, so $I_1^{j_1}$ is convolved with $\psi^{j_2}$ and $\psi^{j_3}$, but $I_1^{j_2}$ is only convolved with $\psi^{j_3}$. 
    }
    \label{fig:ST_form}
\end{figure}

The WST method provides a small set of hierarchical coefficients that serves as multiscale summary statistics of an input field $I_0(\hat{\bm{n}})$. These WST coefficients are obtained through successive convolution of $I_0(\hat{\bm{n}})$ against a family of localized wavelets with different sizes as explained in~\cref{sec:morlet}, followed by a non-linear modulus operation and averaging. The wavelet convolutions decompose the input field into localized features at different spatial scales, while the modulus propagates information to lower frequencies, enabling subsequent wavelet decompositions to capture interactions between structures across multiple scales.  \Cref{fig:ST_form} gives a schematic illustration of how the WST method works.

To obtain the zeroth-order scattering coefficients, we take the spatial average of $I_0(\hat{\bm{n}})$:  
\begin{equation}
    S_0 = \langle I_0 (\hat{\bm{n}}) \rangle.
\end{equation}
Then, in order to obtain the first-order WST coefficient, we perform the following mathematical operations on the initial field $I_0(\hat{\bm{n}})$:
\begin{enumerate}
    \item convolve the initial signal with a set of band-pass filters with different scales $j_1$, denoted by $\psi^{j_1}(\hat{\bm{n}})$: $
    (I_0\ast\psi_j)_{\ell m}
    =
    \psi_j(\ell)(I_0)_{\ell m}
$;
    \item take the modulus of these convolutions, i.e., $I_1^{j_1}(\hat{\bm{n}}) = |I_{0}(\hat{\bm{n}})\ast \psi^{j_1}(\bm{n})|$;
    \item take the global average, i.e., $S_1^{j_1} = \langle I_1^{j_1}(\hat{\bm{n}})\rangle$.
\footnote{In the original formulation of the scattering transform~\cite{Mallat2012GroupScattering}, the final averaging is performed by convolving the modulus field with a low-pass scaling function $\Phi^J(\hat{\bm{n}})$, where the averaging scale $J$ determines the degree of translation invariance. Since the objective of this work is to characterize the global statistical properties of full-sky cosmological fields rather than localized features, we instead compute global scattering coefficients by averaging over the sphere.}
\end{enumerate}

For a Gaussian field, the wavelet response \(W_j(\hat{\mathbf{n}})=I_0\ast\psi^j(\hat{\mathbf{n}})\) is itself Gaussian, and its one-point distribution is therefore completely determined by its variance. Since the expected absolute value of a zero-mean Gaussian variable is proportional to its standard deviation, the first-order coefficient \(S_1^j=\langle|W_j|\rangle\) is determined by the power spectrum. Consequently, the scale dependence of \(S_1\) broadly traces the distribution of power across angular scales~\citep{Mallat2012,Cheng2020ATransform}. For a non-Gaussian field, however, the first absolute moment of \(W_j\) is not determined by its variance alone. Two fields with identical power spectra can therefore have different \(S_1\) coefficients if the one-point distributions of their wavelet responses differ.

The nonlinear modulus is a crucial component of the scattering transform. Without the modulus, successive wavelet convolutions would remain linear operations and would not capture statistical information beyond the power spectrum. This operation removes the oscillatory phase of the wavelet coefficients and generates a slowly varying envelope that can be analyzed by subsequent wavelet convolutions at larger scales. Consequently, higher-order scattering coefficients capture correlations between structures across different spatial scales and are sensitive to non-Gaussian statistical information.

We obtain the second-order coefficients by carrying out these operations on $I_1^{j_1}(\bm{n})$. This can be written in the following form:
\begin{equation}
    I_2^{j_1, j_2}  (\hat{\bm{n}}) = |I_{1}^{j_1}(\hat{\bm{n}})\ast\psi^{j_2}(\hat{\bm{n}})| = ||I_{0}(\hat{\bm{n}})\ast\psi^{j_1}(\hat{\bm{n}})|\ast\psi^{j_2}(\hat{\bm{n}})|,\\
\end{equation}
\begin{equation}
    S_2^{j_1, j_2} = \langle I_2^{j_1,j_2}(\hat{\bm{n}})\rangle.
\end{equation}
The second filter (with scale $j_2$) should have a bigger scale than $j_1$, since the first convolution has already smoothed the field within the scale $j_1$. In general, any higher-order scale should be larger than filters with lower orders, i.e., $j_n>j_{n-1}$. Hence, the total number of wavelet coefficients up to the second order is $1+J+J(J-1)/2$, where $J$ is the maximum scale of our wavelets. For example, for a set of filters with $J=10$, the total number of coefficients up to second order is 56. One could similarly define the third-order coefficients, however, our numerical experiments show that third-order coefficients tend to carry little additional information. 

The second convolution, $I_2^{j_1,j_2}(\hat{\bm{n}})$ picks up structures from $I_1^{j_1}\hat{\bm{n}})$, which represents the local amplitude of structures selected from the original field at scale $j_1$. Thus, the second convolution characterizes cross-scale interactions by measuring how the amplitude of structures at scale \(j_1\) varies over the larger scale $j_2$. Therefore, second-order scattering coefficients are sensitive to higher-order statistics, with a particular close connection to four-point information. 

For a fixed number of wavelet scales \(J\), the computational cost of \texttt{HAWCS} is dominated by repeated spherical-harmonic transforms, which to some maximum multipole $\ell_{\max}$ scale approximately as \(\mathcal{O}(\ell_{\max}^{3})\). The complexity of the code for computing coefficients up to second order is therefore \(\mathcal{O}(J^2 \ell_{\max}^{3})\). This is considerably less expensive than direct general bispectrum and trispectrum calculations, whose reported costs can scale as \(\mathcal{O}(\ell_{\max}^{5})\) to \(\mathcal{O}(\ell_{\max}^{6})\) for the bispectrum~\citep{Komatsu2005,Philcox2023} and up to \(\mathcal{O}(\ell_{\max}^{10})\)~\citep{Philcox2023} for direct trispectrum constructions. Optimized binned or separable estimators can substantially reduce these costs perhaps even to \(\mathcal{O}(\ell_{\max}^{3})\), but they still require evaluating a large number of triangle or quadrilateral configurations~\citep{Komatsu2005,Regan2010}.

\begin{figure}[htbp!]
    \centering
    \includegraphics[width =1 \linewidth]{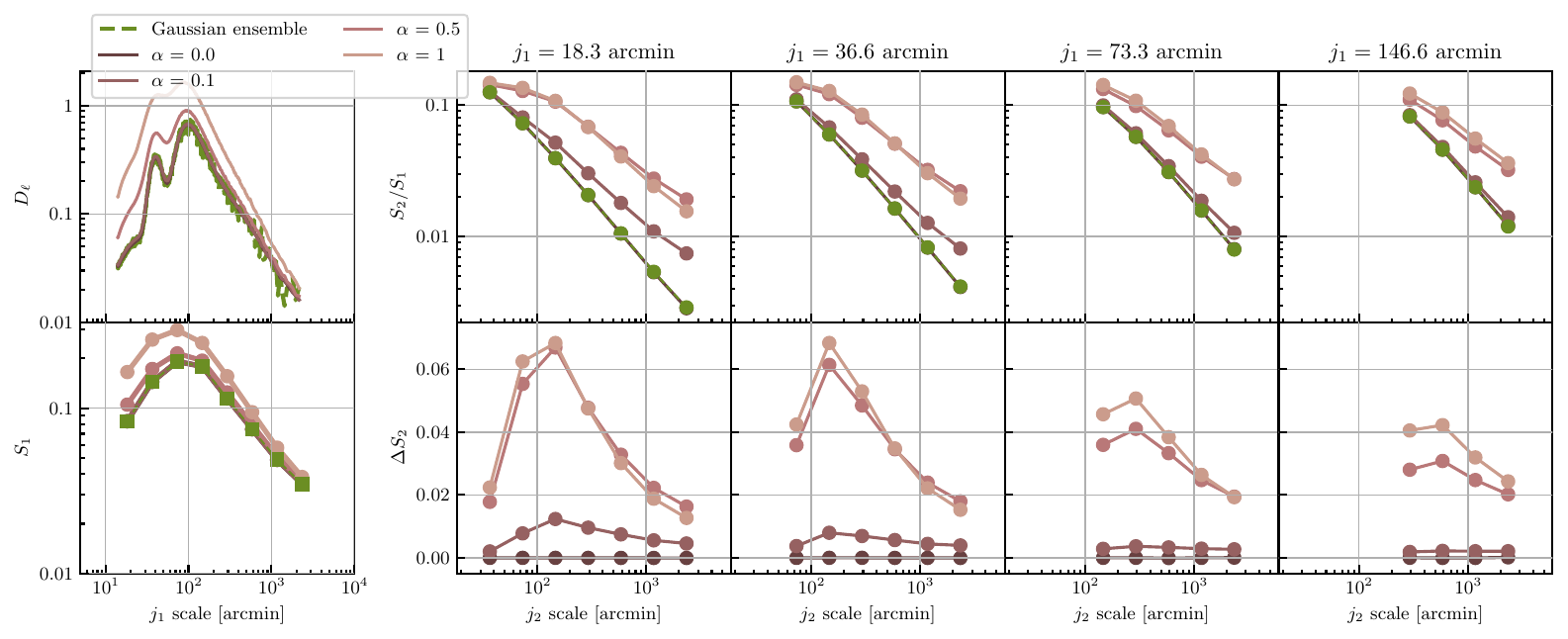}
    \caption{ Test of the response of the scattering coefficients to quadratic non-Gaussianity. The left panels show the angular power spectrum \(C_\ell\) and first-order coefficients, \(S_1\). The top right panels show the normalized second-order coefficients, \(S_2(j_1,j_2)/S_1(j_1)\), as functions of the second-layer scale, \(j_2\), for different first-layer scales, \(j_1\).  The bottom right panels show the quantiry $\Delta S_2$ for these maps. The Gaussian reference agrees with the \(\alpha=0\) case and they show an exponential decrease with scale, while increasing \(\alpha\) produces a progressively stronger scale-dependent response in the second-order coefficients. These diagrams confirm that second-order coefficients are very effective in distinguishing between Gaussian and non-Gaussian fields. The error bars, estimated from the scatter of 100 realizations, are smaller than the plotted markers. The quadratic fields are therefore clearly distinguishable from the Gaussian expectation.
    }
    \label{fig:gaus_sim}
\end{figure}

For a Gaussian field, these higher-order statistics are completely determined by the 2-point function and the resulting $S_2$ coefficients effectively measure the overlap between two wavelets and so they exhibit a characteristic rapid decay with increasing scale, as shown in~\cref{fig:gaus_sim}. However, a non-Gaussian field may contain an additional connected four-point contribution. Therefore, fields with identical power spectra can exhibit different $S_2$
 coefficients owing to differences in their cross-scale amplitude modulation and higher-order spatial structure. This makes $S_2$
 effective for distinguishing Gaussian and non-Gaussian fields. 

The $S_2$ coefficients are affected by observational effects such as beam smoothing, instrumental noise, and masking, which can suppress existing structures or introduce additional scale-dependent features. 
Their amplitudes also depend on the overall normalization of the angular power spectrum because both \(S_1\) and \(S_2\) scale linearly under a global rescaling of the field. To account for these effects and isolate non-Gaussianity, we generate Gaussian control maps with matched angular power spectra and subject them to the same beam, noise, and masking. We define the excess $S_2$ parameter as
\begin{equation}
\Delta S_2(j_1,j_2)
=
\frac {S_2^{\mathrm{field}}(j_1,j_2)}{S_1^{\mathrm{field}}(j_1)}
-
\left\langle \frac{S_2^{\mathrm{G}}(j_1,j_2)}{S_1^{\mathrm{G}}(j_1)}\right\rangle ,
\end{equation}
where the average is taken over the ensemble of matched Gaussian control maps. We find that this difference is very robust to beam, noise, and masking effects.

To validate our spherical WST implementation, we construct a controlled family of fields with a tunable level of non-Gaussianity and compare their scattering coefficients with those of an ensemble of Gaussian control maps with the same angular power spectrum. We first generate a Gaussian field \(u\) from an arbitrary input power spectrum and then apply the local quadratic transformation
$
x_\alpha
=
u+\alpha \, u^2.
$
The parameter \(\alpha\) controls the amplitude of the quadratic (non-Gaussian) term \(u^2\). For \(\alpha=0\), \(x_\alpha=u\) and the field is Gaussian. Increasing \(\alpha\) introduces progressively stronger non-Gaussianity through the quadratic term, providing a simple controlled test of the response of the second-order scattering coefficients. To capture the scatter of the coefficients, for each $\alpha$ value, we generate 100 realizations. As can be seen in \cref{fig:gaus_sim}, the power spectra of the simulations are very similar and the $S_1$ coefficients roughly follow the shape of the power spectra, as expected because they both primarily probe scale-dependent power. The $S_2$ coefficients however, show the coupling with structures at different scales and exhibit a totally different structure. Additionally, as anticipated, the deviation from Gaussian shape increases with $\alpha$. The quantity $\Delta S_2$ shows the deviation of non-Gaussian maps from one another more clearly. 

\section{Scientific Applications}
\label{section:results}
\subsection{Comparing different simulations of the SZ map}

The thermal Sunyaev-Zeldovic (tSZ) signal arises from inverse Compton scattering of cosmic microwave background photons by hot, ionized gas, primarily in galaxy groups and clusters. Since the tSZ signal is dominated by a sparse population of massive halos and traces their highly structured gas distribution, the resulting Compton $y$ field is strongly non-Gaussian. Its morphology depends on both the cosmological halo population and the astrophysical processes governing the intracluster gas. Consequently, realistic simulations must reproduce not only the tSZ power spectrum, but also the higher-order and cross-scale information associated with the abundance, spatial distribution, internal structure, and dynamical state of halos. It is therefore useful to use \texttt{HAWCS} to evaluate how successful simulations are at replicating the non-Gaussian structure in the tSZ data. 

We use the \textit{Planck} $y$ map. \textit{Planck} observes the sky at multiple frequencies, each frequency map containing a mixture of the tSZ signal, primary CMB emission, Galactic foregrounds, extragalactic sources, and instrumental noise. Component-separation methods combine these frequency maps by exploiting the distinct frequency dependence of each emission component, with the aim of isolating the signal of interest while suppressing contaminants. The $y$ maps used here were reconstructed by the \textit{Planck} Collaboration using the Modified Internal Linear Combination Algorithm (\texttt{MILCA}) and Needlet Internal Linear Combination (\texttt{NILC})~\cite{planck2014-a28}. 

We choose to examine four simulations of the SZ sky constructed using distinct numerical and astrophysical prescriptions: \texttt{Websky}~\cite{Stein2018TheValidation,Stein2020TheSimulations}, generates a large-volume dark-matter light cone and assigns gas-pressure profiles to halos to produce a full-sky $y$ map. The rescaled Sehgal et al.\ simulation used in the Simons Observatory forecasts~\cite{Sehgel2010} similarly combines an $N$-body matter distribution with a semi-analytic model for gas in halos and the intergalactic medium. \texttt{Magneticum}~\cite{Coulton2021EffectsStudies}, in contrast, is a hydrodynamical simulation in which the dark matter and baryonic gas are evolved together, and the $y$ map is obtained by projecting the simulated electron pressure along the line of sight. Finally, \texttt{HalfDome}~\cite{Bayer2025HalfDome} uses large-volume FastPM $N$-body simulations to generate full-sky light cones, from which $y$ maps are constructed by assigning empirical electron-pressure profiles to the resolved halo population; its multiple fixed-cosmology realizations additionally allow us to estimate realization-to-realization scatter.

\begin{figure}[htbp!]
    \centering
    \includegraphics[width =1 \linewidth]{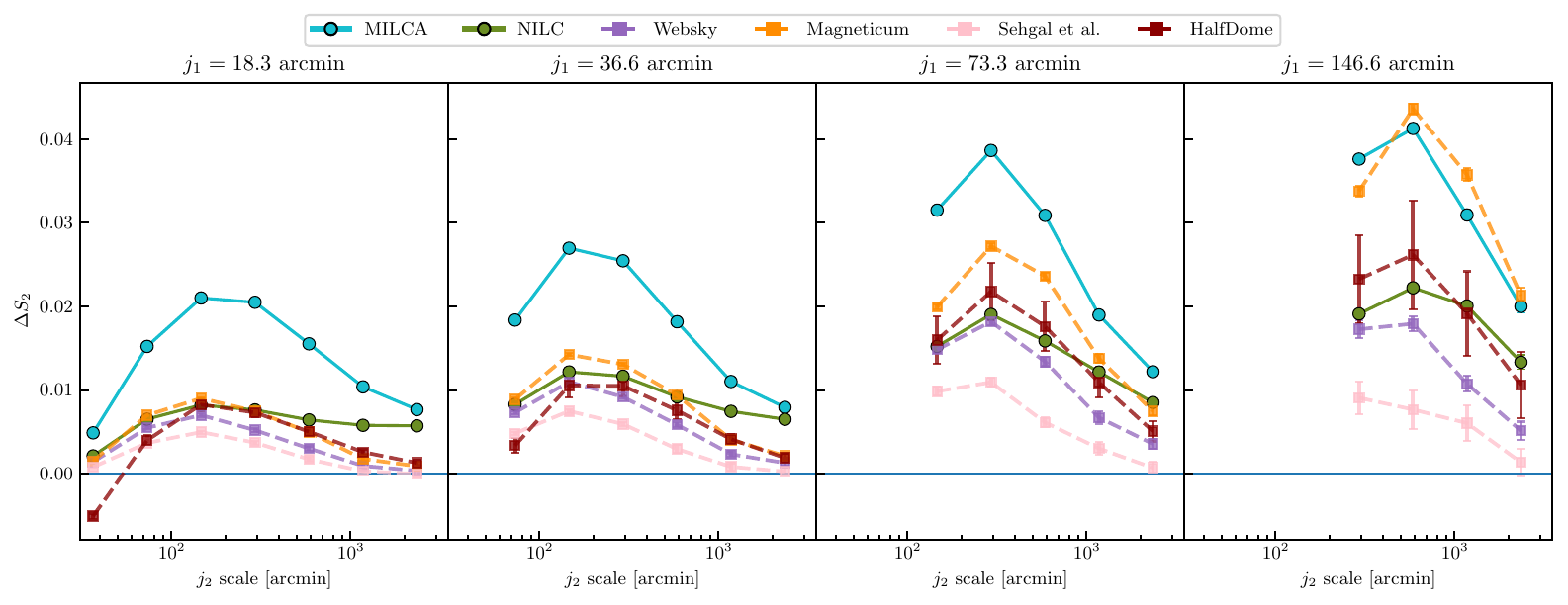}
    \caption{ $\Delta S_2$ coefficient for tSZ maps for \textit{Planck} \texttt{MILCA} and \texttt{NILC} component-separation methods and one realization from \texttt{Websky}, \texttt{Magneticum}, and Sehgal et al.\ simulations in addition to 11 realizations of the \texttt{HalfDome} simulations. The error bars on the \texttt{Websky}, Magneticum, and Sehgal et al.\ simulations are from single realizations, and therefore they do not account for the realization-to-realization variance of the simulations; they only indicate the significance of the departure of these simulations from Gaussian null assumption. The error bara on \texttt{HalfDome}, however, comes from the scatter between the different realizations. The \texttt{MILCA} and \texttt{NILC} measurements differ appreciably, with \texttt{MILCA} generally exhibiting a larger non-Gaussian excess, particularly at intermediate angular scales. The simulations are more consistent with \texttt{NILC} in amplitude. This comparison demonstrates that the inferred tSZ non-Gaussianity depends significantly on the component-separation method and that the \texttt{MILCA}--\texttt{NILC} difference is comparable to, or larger than, the variation among the simulations. }
    \label{fig:sz}
\end{figure}

The results in \Cref{fig:sz} show a clear  discrepancy between \texttt{MILCA} and \texttt{NILC}, despite both maps being reconstructed from the same \textit{Planck} frequency data. This difference likely reflects sensitivity to the component-separation procedure, including differing residual foreground contamination, scale-dependent filtering, and spatially varying noise or scan-depth effects. The simulations also yield distinct \(\Delta S_2\) patterns, indicating that their higher-order cross-scale structure is not identical even when their power spectra are broadly similar. In overall amplitude, however, the simulated maps are generally more consistent with \texttt{NILC} than with \texttt{MILCA}; none fully reproduces the detailed dependence on the two wavelet scales seen in either reconstruction. The error bars for \texttt{HalfDome} show the scatter among 11 realizations with fixed cosmology. Since this scatter is smaller than several of the differences among simulations and between the two \textit{Planck} maps, these discrepancies cannot be attributed solely to cosmic variance. For the other simulations, only a single realization was available, so their realization-to-realization uncertainty is not shown. Their error bars represent the scatter of the matched-Gaussian ensemble. These results indicate that the WST captures information not constrained by the tSZ power spectrum. It responds to the spatial organization of halo pressure profiles and their environments, while also being sensitive to differences in residual foregrounds and filtering introduced during component separation.

\subsection{CMB lensing}

Gravitational lensing by large-scale structure remaps the primary CMB temperature field (see Ref.~\cite{LewisChallinor} for a review) according to
\begin{equation}
T^{\mathrm{len}}(\hat{\boldsymbol n})
=
T^{\mathrm{unl}}\!\left(\hat{\boldsymbol n}+\nabla\phi(\hat{\boldsymbol n})\right),
\end{equation}
where \(\phi\) is the lensing potential and \(\nabla\phi\) is the corresponding deflection field. To first order in the deflection, this remapping can be expanded as
\begin{equation}
T^{\mathrm{len}}(\hat{\boldsymbol n})
\simeq
T^{\mathrm{unl}}(\hat{\boldsymbol n})
+
\nabla\phi(\hat{\boldsymbol n})
\boldsymbol{\cdot}
\nabla T^{\mathrm{unl}}(\hat{\boldsymbol n}).
\end{equation}
Although the unlensed CMB is well approximated as a Gaussian random field, this position-dependent deflection couples harmonic modes and generates non-Gaussian correlations, most notably a connected four-point function. It couples CMB temperature modes \(\ell\) and \(\ell'\) through a lensing mode \(L\sim\ell-\ell'\). 

The WST provides a natural way to characterize this lensing-induced structure.
The first-layer wavelet selects small-scale CMB fluctuations and even though a nearly uniform deflection only translates these fluctuations and has little effect on the WST (which is translation-invariant), their local amplitudes are instead modulated by spatial variations in the deflection field, described by the convergence and shear. Since the convergence satisfies
\begin{equation}
\kappa_{LM}
=
\frac{L(L+1)}{2}\phi_{LM},
\qquad
C_L^{\kappa\kappa}
=
\frac{[L(L+1)]^2}{4}C_L^{\phi\phi},
\end{equation}
the scale dependence of \(\Delta S_2\) is expected to be more closely connected to the convergence or shear power than to \(C_L^{\phi\phi}\) itself. Schematically,

\begin{equation}
\Delta S_2(j_1,j_2)
\sim
\sum_{\ell,L}
(L (L+1))^2 \,
C_L^{\phi\phi}\,
\mathcal{R}_{j_1j_2}(\ell,L, C_\ell ^{CMb}),
\end{equation}

\begin{figure}[htbp!]
    \centering
    \includegraphics[width =1 \linewidth]{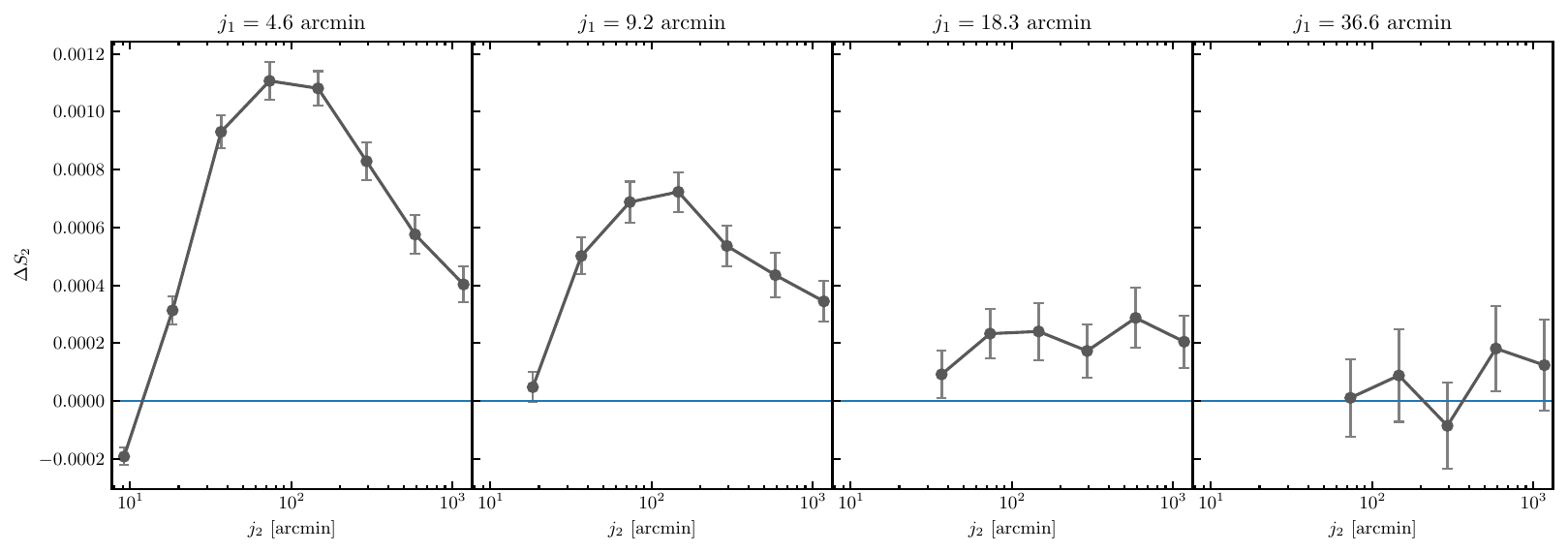}
    \caption{Excess second-order scattering coefficients of a simulated \textit{Planck} lensed CMB temperature map relative to the mean of matched-spectrum Gaussian realizations. The lensing-induced excess is strongest for few arcmin scales and decreases as the first-layer scale increases, indicating that lensing primarily modifies the cross-scale organization of small-scale CMB fluctuations. 
    }
    \label{fig:lensing2}
\end{figure}

\Cref{fig:lensing2} shows $\Delta S_2$ for a simulated \textit{Planck} lensed CMB map. The strongest departures occur when the first-layer wavelet selects the smallest CMB scales. Several coefficients in the $(j_1=4.6)$ arcmin panel differ substantially from the Gaussian mean. These values represent pointwise deviations; because neighboring coefficients are correlated, they should not be interpreted as independent or as a joint $(10\sigma)$ detection. 
The lensing-induced residual is largest when the first wavelet selects the smallest CMB scales and decreases progressively as \(j_1\) increases. This behavior indicates that lensing primarily changes the large-scale spatial organization of small-scale CMB fluctuations. 

For fixed \(j_1\), the residual reaches a maximum at an intermediate second-layer scale, suggesting a characteristic scale for the modulation of the first-layer amplitudes. 
The lensing multipole \(L\) labels the mode of the deflection field that couples CMB multipoles. Consequently, \(C_L^{\kappa\kappa}\) controls the scale dependence and amplitude of the leading lensing-induced CMB four-point correlations. The maximum of \(\Delta S_2\) may therefore reflect the lensing scales that most strongly couple the CMB modes selected by the two wavelets. The strong response to lensing-induced cross-scale coupling suggests that these coefficients could provide a useful diagnostic to test if a delensed CMB map has really been delensed.  It has the advantage of lensing-specific estimators of being very quick to evaluate using \texttt{HAWCS}.  Additionally, after delensing after calibration with simulations, these coefficient could provide a complementary summary statistic for parameters controlling the lensing amplitude.

\section{Discussion}
In this work, we have described our implementation of the spherical scattering transform in the form of the \texttt{HAWCS} code, and explained how it can be useful as a summary statistic for detecting non-Gaussianity in sky maps. The computational cost of this code is $\mathcal{O}(J^2\ell_{\max}^{3})$, which is much smaller than general bispectrum and trispectrum computation codes. However, the main
computational advantage of \texttt{HAWCS} is its compact, model-independent representation of non-Gaussian information using relatively few scale-indexed coefficients, without explicitly
enumerating all triangle or quadrilateral configurations.

This method is especially useful for CMB secondary maps, since they trace the non-Gaussian signatures coming from the large-scale structure of the Universe. The information extracted from these maps can be useful in constraining models and parameters, for investigating the quality of noise removal, delensing or dust decontamination, and also for simulating new realizations of these secondary signals. 

The first-order coefficients, $(S_1)$, primarily characterize the scale-dependent amplitude of fluctuations and are therefore closely related to the angular power spectrum. By contrast, the second-order coefficients, $(S_2)$, quantify spatial variations in the amplitudes of the first-layer wavelet coefficients and encode interactions between structures at different scales. Comparisons with Gaussian fields matched in angular power spectrum show that $(S_2)$ can distinguish fields that have identical 2-point statistics but different higher-order structure.

We presented two scientific cases in this paper. First, we investigated the thermal Sunyaev--Zeldovich maps. We compared two \textit{Planck} component-separation maps, \texttt{MILCA} and \texttt{NILC}, with four different simulations. Even though \texttt{MILCA} and \texttt{NILC} are reconstructed from the same data, differences in their component-separation procedures lead to appreciably different non-Gaussian signatures. In particular, \texttt{MILCA} generally exhibits a larger non-Gaussian excess than \texttt{NILC}, while the simulations are more consistent with \texttt{NILC} in overall amplitude. This result demonstrates that higher-order statistics of reconstructed tSZ maps can be sensitive to residual foreground contamination and map-making choices, and that agreement with the tSZ power spectrum alone does not ensure that simulations reproduce the observed cross-scale morphology of the field.

For the second application, we studied gravitational lensing. Weak lensing modulates small-scale CMB fluctuations through coherent deflections and therefore couples angular scales and produces non-Gaussian structure, even though the unlensed CMB is Gaussian. We found that \texttt{HAWCS} detects this lensing-induced cross-scale structure through a clear excess in the second-order coefficients relative to Gaussian fields with matched power spectra. This demonstrates that the method is sensitive to the higher-order information generated by lensing and can rapidly distinguish lensed fields from Gaussian realizations without explicitly reconstructing the lensing potential.

\section*{Acknowledgments}
This work was supported by the Natural Sciences and Engineering Research Council of Canada and the Canadian Space Agency. 
Based on observations obtained with \textit{Planck} (\url{http://www.esa.int/Planck}), an ESA science mission with instruments and contributions directly funded by ESA Member States, NASA, and Canada. 

\section*{Data availability}

\bibliographystyle{unsrturl}

\bibliography{Biblio}
\appendix

\section{Code Implementation}

The spherical scattering analysis is implemented in the Python package
\texttt{HAWCS}. The package is organized as a small modular library, with the
main user interface provided by the class \texttt{ScatteringSph}. This class
wraps the filter construction, harmonic-space convolutions, coefficient
calculation, and optional storage of intermediate maps. A typical initialization is
\begin{verbatim}
from hawcs import ScatteringSph

scattering = ScatteringSph(
    nside=nside,
    J=J,
    order=order,
    lmax=lmax,
    norm="L1",
)
\end{verbatim}
where \texttt{nside} sets the \texttt{HEALPix}~\cite{HEALPix} resolution, \texttt{J} sets the number of
wavelet scales, \texttt{order} sets the maximum scattering order, and
\texttt{lmax} controls the maximum spherical harmonic multipole used in the
\texttt{Healpy} transforms. If \texttt{lmax} is not supplied, the code defaults to
\texttt{3*nside - 1}. The argument \texttt{norm} controls the wavelet
normalization and may be set to \texttt{"L1"}, \texttt{"L2"}, or \texttt{None}.

The filters are generated internally by the functions in
\texttt{hawcs.wavelets}. The package constructs Morlet wavelets and Gaussian
smoothing filters on an angular grid and converts them to harmonic space. The
high-level class caches these harmonic filters so that, when the same
\texttt{ScatteringSph} object is reused for several maps, the filters do not
need to be rebuilt each time.

The coefficients are computed by calling the scattering object directly:
\begin{verbatim}
coefficients = scattering(
    hmap,
    mask=mask,
    keep_maps=False,
)
\end{verbatim}
The input \texttt{hmap} is a \text{HEALPix} map stored as a 1-dimensional array. The
optional \texttt{mask} argument is used when averaging the final coefficient
maps. It can be either a binary mask or an apodized mask with values between
zero and one. Pixels with mask value zero are excluded from the average, while
pixels with intermediate values are included with the corresponding weight. If
the input map is a \texttt{NumPy} or \texttt{Healpy} masked array, the masked pixels are also
excluded automatically.

The output is a nested dictionary. The first level labels the scattering order,
and the second level labels the scale path:
\begin{verbatim}
coefficients["S0"][()]
coefficients["S1"][(j1,)]
coefficients["S2"][(j1, j2)]
coefficients["S3"][(j1, j2, j3)]
\end{verbatim}
This structure was chosen so that each coefficient can be accessed explicitly by
its scattering order and by the wavelet scales used to compute it. For example,
\texttt{coefficients["S2"][(1, 3)]} returns the second-order coefficient
corresponding to the scale path \texttt{j1=1, j2=3}.

The keyword \texttt{keep\_maps} controls whether intermediate modulus maps are
stored. When \texttt{keep\_maps=True}, the intermediate maps are saved in
\texttt{scattering.intermediate\_maps\_} and can be accessed using helper
methods such as
\begin{verbatim}
I1 = scattering.get_i1(j)
I2 = scattering.get_i2(j1, j2)
\end{verbatim}
This is useful for diagnostic plots or for inspecting the intermediate
scattering maps. For large \texttt{HEALPix} maps, however, storing these arrays can
require substantial memory. Therefore, for production runs we usually set
\texttt{keep\_maps=False}. In this mode the code computes one scale path at a
time and discards intermediate maps as soon as they are no longer needed.

The package also provides lower-level functions in separate modules. The module \texttt{hawcs.wavelets} contains the Gabor, Morlet, Gaussian, and filter-bank
construction functions. In addition to using these filters internally,
\texttt{ScatteringSph} can return the filter bank directly in either real space
or harmonic space. For example,
\begin{verbatim}
filters_real = scattering.filter_bank(space="real")
filters_ell = scattering.filter_bank(space="harmonic")
\end{verbatim}
returns dictionaries containing the Morlet wavelets in \texttt{filters["psi"]}
and the Gaussian filters in \texttt{filters["phi"]}. The real-space filters are
sampled as functions of angular separation, while the harmonic-space filters are
stored as functions of multipole $\ell$ and are the filters used in the \texttt{Healpy}
convolutions. 

 \end{document}